\documentclass[aps,prl,showpacs,superscriptaddress,twocolumn]{revtex4}

\usepackage{hyperref}
\usepackage{bbm}
\usepackage{graphicx}
\usepackage{amsmath}
\usepackage{amssymb}
\usepackage{mathscinet}
\usepackage{amsfonts,amsmath}
\usepackage{color} 

\usepackage{xparse}
\usepackage{tikz}
\usetikzlibrary{matrix,backgrounds}
\pgfdeclarelayer{myback}
\pgfsetlayers{myback,background,main}

\usepackage{marvosym}
\usepackage{wasysym}

\tikzset{mycolor/.style = {line width=1bp,color=#1}}%
\tikzset{myfillcolor/.style = {draw,fill=#1}}%

\NewDocumentCommand{\highlight}{O{blue!40} m m}{%
\draw[mycolor=#1] (#2.north west)rectangle (#3.south east);
}

\NewDocumentCommand{\fhighlight}{O{blue!40} m m}{%
\draw[myfillcolor=#1] (#2.north west)rectangle (#3.south east);
}

\def\one{{\mathchoice {\rm 1\mskip-4mu l} {\rm 1\mskip-4mu l} {\rm
1\mskip-4.5mu l} {\rm 1\mskip-5mu l}}}

\usepackage[matrix,frame,arrow]{xypic}

\begin{document}

\title{Achievable polarization for Heat-Bath Algorithmic Cooling}

\author{Nayeli A. Rodr\'i­guez-Briones}

\affiliation{Institute for Quantum Computing, University of Waterloo, Waterloo, Ontario, N2L 3G1, Canada}
\affiliation{Department of Physics \& Astronomy, University of Waterloo, Waterloo, Ontario, N2L 3G1, Canada}

\author{Raymond Laflamme}

\affiliation{Institute for Quantum Computing, University of Waterloo, Waterloo, Ontario, N2L 3G1, Canada}
\affiliation{Department of Physics \& Astronomy, University of Waterloo, Waterloo, Ontario, N2L 3G1, Canada}
\affiliation{Perimeter Institute for Theoretical Physics, 31 Caroline Street North, Waterloo, Ontario, N2L 2Y5, Canada}
\affiliation{Canadian Institute for Advanced Research, Toronto, Ontario M5G 1Z8, Canada}

\date{\today}

\begin{abstract}
Pure quantum states play a central role in applications of quantum information, both as initial states for quantum algorithms and as resources for quantum error correction. Preparation of highly pure states that satisfy the threshold for quantum error correction remains a challenge, not only for ensemble implementations like NMR or ESR but also for other technologies. Heat-Bath Algorithmic Cooling is a method to increase the purity of a set of qubits coupled to a bath. We investigated the achievable polarization by analysing the limit when no more entropy can be extracted from the system. In particular we give an analytic form for the maximum polarization achievable for the case when the initial state of the qubits is totally mixed, and the corresponding steady state of the whole system. It is however possible to reach higher polarization while starting with certain states, thus our result provides an achievable bound. We also give the number of steps needed to get a specific required polarization.

\end{abstract}

\pacs{03.67.Pp}

\maketitle

\section{Introduction}
Purification of quantum states is essential for applications of quantum information science, not only for many quantum algorithms but also as a resource for quantum error correction. The need to find a scalable way to reach approximate pure states is a challenge for many quantum computation modalities, especially the ones that relies on ensembles such as NMR or ESR~\cite{Ladd:2010uq}.

A potential solution is algorithmic cooling (AC), a protocol which purifies qubits by removing entropy of a subset of them, at the expense of increasing the entropy of others~\cite{sorensen:qc1990a, sorensen1991entropy}. An explicit way to implement this idea in ensemble quantum computers was given by Schulman et al. \cite{schulman:qc1998a}. They showed that it is possible to reach polarization of order unity using only a number of qubits which is polynomial in the initial polarization. This idea was improved by adding contact with a heat-bath to extract entropy from the system \cite{boykin2002algorithmic}, a process known as Heat-Bath Algorithmic Cooling (HBAC). Based on this work, many cooling algorithms have been designed
\cite{fernandez2004algorithmic,moussa:2005,schulman2005physical,elias2006optimal,schulman2007physical,elias2011semioptimal}. HBAC is not only of theoretical interest, experiments have already demonstrated an improvement in polarization using this protocol with a few qubits \cite{fernandez2005paramagnetic,baugh2005experimental,Ryan:2008qf,elias2011heat,Brassard:2014fk,brassard2014prospects,park2014hyperfine}, where a few rounds of HBAC were reached; and some studies have even included the impact of noise~\cite{kaye:2007}.

Through numerical simulations, Moussa \cite{moussa:2005} and Schulman et al. \cite{schulman2005physical} observed that if the polarization of the bath ($\epsilon_b$) is much smaller than $2^{-n}$, where $n$ is the number of qubits used, the asymptotic polarization reached will be $\sim 2^{n-2} \epsilon_b$; but when $\epsilon_b$ is greater than $2^{-n}$, a polarization of order one can be reached. Inspired also by the work of Patange \cite{Patange:2013}, who investigated the use of algorithmic cooling on spins bigger than $\frac{1}{2}$ (using NV center where the defect has an effective spin 1), we investigate the case of cooling a qubit using a general spin $l$, and extra qubits which get contact with a bath. We found the asymptotic limit by solving the evolution equation with the results supported by numerical simulation \cite{moussa:2005}. A proof has been reported by Raeisi and Mosca \cite{raeisi2015asymptotic}.

In this paper we give the analytic result for the asymptotic polarization that can be reached when the initial state of the quantum computer is in the totally mixed state. This gives an achievable bound as we can always efficiently turn a state into the maximally mixed one, while some other initial states do lead to higher polarizations. We recover the limit of low polarization observed by Moussa and Schulman et al. We also show how a polarization of order one can be reached as a function of the number of qubits. We compare the Schulman's upper bound of the maximum probability of any basis state \cite{schulman2007physical} with our analytical bound. Finally we give the number of rounds of compression/cooling needed to get certain polarization.


\begin{figure}[h]
\centering
\includegraphics[width=0.45\textwidth]{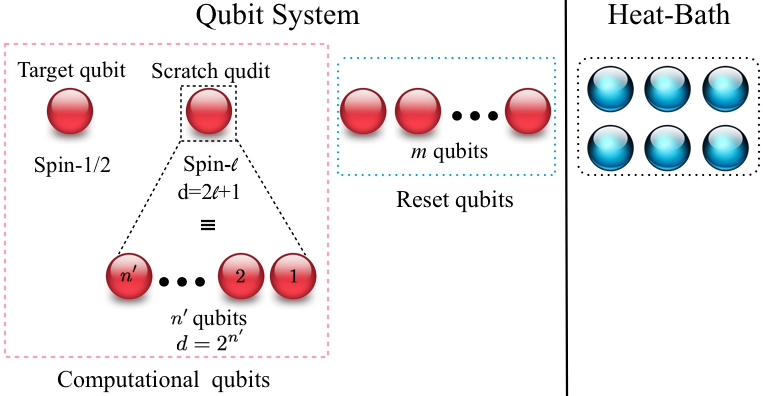}
\caption{
HBAC can cool the target qubit by compressing entropy into $m$ reset qubits and a $d-$dimensional spin$-l$ (or a string qubits of Hilbert space of dimension $d$); then, HBAC pumps entropy from the qubit system into a heat-bath by refreshing the $m$ reset qubits.}
\label{fig:system}
\end{figure}

HBAC purifies qubits by applying alternating rounds of entropy compression and pumping entropy into a thermal bath of partially polarized qubits, as explained below.

The system consists of a string of qubits: one qubit (spin$-1/2$, also called the target qubit) which is going to be cooled; one qudit (called the scratch system, which can be a spin$-l$ or a string of qubits) which aids in the entropy compression; and $m$ reset qubits that can be brought into thermal contact with a heat-bath of polarization $\epsilon_b$. Having the spin$-l$ is equivalent to having $n'$ qubits if the dimension of their Hilbert spaces is the same, i.e. if $d=2l+1=2^{n'}$. We will also refer to the target qubit and the scratch qudit as the computational qubits (Fig. \ref{fig:system}).

The idea of HBAC is to first re-distribute the entropy among the string of qubits applying an entropy compression operation $U$. This operation concentrates the entropy in the reset qubits of the system by extracting entropy of the computational qubits. This process results in the cooling of the computational qubits while warming the reset qubits. The second step is to refresh the system using the heat-bath for removing entropy.

For our study, we used a HBAC algorithm called the Partner Pairing Algorithm (PPA), which was invented by Schulman et al. \citep{schulman2005physical}. This protocol gives the optimal physical cooling of HBAC, in terms of entropy extraction, under the assumption that the refresh step rethermalizes the reset qubits with the heat-bath \citep{schulman2005physical,schulman2007physical}. In the PPA, the entropy compression operation, $U$, makes a descending sort of the diagonal elements of the system's density matrix. In the refresh step, the $m$ reset qubits are brought into thermal contact with the bath to be refreshed. This step is equivalent to tracing-over the reset qubits, and replacing them with qubits from the heat-bath, cooling the qubit system. We also assume that the heat bath has large heat capacity and that the action of qubit-bath interaction on the bath temperature is negligible.

The total effect of applying these two steps on a system with state $\rho$ can be expressed as follows:
\begin{align}
&\rho\xrightarrow{Compression} \rho'= U\rho U^{\dagger}, \\
& \ \ \ \ \rho'  \xrightarrow{Refresh}  \rho''=\mathrm{Tr}_{m_{qubits}}\left(\rho'\right)\otimes\rho_{\epsilon_b}^{\otimes m},
\label{eq:refresh}
\end{align}
where $\rho_{\epsilon_b}= \frac{1}{2} \begin{pmatrix}
 1+\epsilon_b& 0\\
 0 & 1-\epsilon_b\\
 \end{pmatrix}$ is the state of a qubit from the bath, and $\epsilon_b$ is the heat-bath polarization (some authors, such as Schulman et al. \cite{schulman2007physical}, use $\epsilon={\rm arctanh} \epsilon_b$ as polarization).

An interesting question is what is the asymptotic achievable cooling with this method, and how many iterations of the HBAC-steps would be needed to obtain a certain cooling, i.e. a certain value of polarization.

\section{Cooling Limit}
The cooling limit corresponds to the moment at which it is not possible to continue extracting entropy from the system, i.e. \textit{when the state of the qubit system is not changed by the compression and refresh steps}. The system achieves this limit asymptotically, converging to a steady state where the following condition holds: 
\begin{equation}
\label{eq:cond}
\rho=\rho''.
\end{equation}
The state of the computational qubits, $\rho_{com}=\mathrm{Tr}_{m_{qubits}}\left(\rho\right)$, can be expressed as
\begin{equation}
diag(\rho_{com})=\left(A_1, A_2, A_3, ..., A_{2d}\right),
\end{equation}
where $diag(\rho)$ is the vector of the diagonal elements of $\rho$. From this and eq.(\ref{eq:refresh}), the state of the qubit system after a HBAC iteration will be described by
\begin{equation}
diag(\rho'')=\left(A_1, A_2, ..., A_{2d}\right)\otimes\frac{1}{2^m}\left(1+\epsilon,1-\epsilon\right)^{\otimes m}.
\end{equation}


In the cooling limit there is no operation that can compress any further the entropy of the computational qubits, or equivalently, the diagonal elements of $\rho''$ are already sorted in decreasing order. This will happen when we have the condition
\begin{equation}
\label{eq:cond1}
A_{i}\left(1-\epsilon_b\right)^m \geq A_{i+1}\left(1+\epsilon_b\right)^m,
\end{equation}
 for $i=1, 2, 3,...,2d-1$, (see Supplemental Material for details). When this equation is satisfied, the entropy of the reset qubits will not increase anymore after compression and thus contact with the bath will not cool them. Thus, HBAC iterations will not modify the state anymore, leading to (\ref{eq:cond}).

\section{Maximally mixed initial state}

If we start with a maximally mixed state, it is possible to show that (see supplemental material, a proof can be found in \cite{raeisi2015asymptotic}) 
\begin{equation}
A_{i}^t\left(1-\epsilon_b\right)^m \leq A_{i+1}^t\left(1+\epsilon_b\right)^m,
\label{eq:inequal}
\end{equation}
where $t$ labels the number of HBAC iterations. This is true for the initial step, as $A_i=\frac{1}{2d}$ for all $i$ at $t=0$, but it turns out that it remains true for all subsequent iterations. 

It is also possible to show that at each step the polarization of the target qubit never decreases, while the entropy of the reset qubits always increases beyond the one from the bath at each entropy compression step. Thus, the reset qubits always pump entropy out of the system into the bath, converging to a limit.

Comparing eq.(\ref{eq:cond1}) and eq.(\ref{eq:inequal}) indicates that the asymptotic state 
of the computational qubits can only go towards the equality
\begin{equation}
\label{eq:cond2}
A^{\infty}_{i}\left(1-\epsilon_b\right)^m=A^{\infty}_{i+1}\left(1+\epsilon_b\right)^m,
\end{equation}
 for all $i=1,2,3,...,2d-1$.
 
From (\ref{eq:cond2}) and the property ${\rm Tr}\left(\rho_{com}\right)=1$, it is possible to find $A^{\infty}_i=\frac{1-Q}{1-Q^{2d}}Q^{i-1}$, where $Q=\left(\frac{1-\epsilon_b}{1+\epsilon_b}\right)^m$. This result gives the exact solution of the steady state of the computational qubits, $\widetilde\rho_{com}$, for all values of the bath polarization:
 \begin{equation}
\label{eq:steadystate}
diag\left(\widetilde\rho_{com}\right)=A^{\infty}_1\left(1, Q, Q^2,..., Q^{2d-1}\right).
\end{equation}
See Supplemental Material for details.


\subsection{Asymptotic Polarization}

From the steady state (eq. (\ref{eq:steadystate})), the asymptotic polarization of the target qubit is
\begin{equation}
\label{eq:maxpol}
\epsilon_{\one}^{\infty}=\frac{\left(1+\epsilon_b\right)^{md}-\left(1-\epsilon_b\right)^{md}}{\left(1+\epsilon_b\right)^{md}+\left(1-\epsilon_b\right)^{md}}.
\end{equation}
The corresponding temperature of the target qubit will be $T_{steady}=\frac{1}{md}T_b \frac{\Delta E_t}{\Delta E_r}$ ($d=2^{n'}$ when the scratch qudit is a string of $n'$ qubits), here $T_b$ is the temperature of the bath, and $\Delta E_t$ and $\Delta E_r$ are the energy gaps between the two energy levels of the target qubit, and the reset qubits, respectively. Our results agree with the third law of thermodynamics \citep{levy2012quantum,masanes2014derivation}.

For the case of using a string of qubits as the scratch qudit, the maximum achievable polarization of the $j^{th}$ qubit will be $ \epsilon^{\left(j\right)}_{max}=\frac{\left(1+\epsilon_b\right)^{m2^{j-1}}-\left(1-\epsilon_b\right)^{m2^{j-1}}}{\left(1+\epsilon_b\right)^{m2^{j-1}}+\left(1-\epsilon_b\right)^{m2^{j-1}}}$ (numbered from right to left, Fig. \ref{fig:system}).

\begin{figure}[h] 
\center{\includegraphics[width=0.9\linewidth]{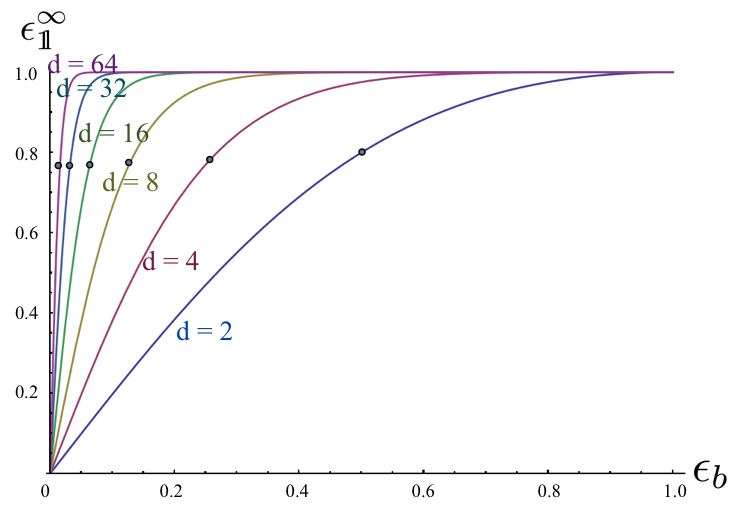}}
\caption{Asymptotic achievable polarization for the target qubit. This polarization increases double exponentially in the number of qubits as the scratch qudit, $n'$. The dots are located at the point of $\epsilon_{\one}^{\infty}$ which corresponds to the $\epsilon_b=\frac{1}{md}$, where the transition can be observed, for $d=2$, $4$, $8$, $16$, $32$, and $64$, and $m=1$. (For $\epsilon_b$ smaller than that value, $\epsilon_{\one}^{\infty}$ is linear in $\epsilon_b$.)}
\label{fig:polmax_d}
\end{figure}

In the limit for low bath polarization, $\epsilon_b<<1/md$, the achievable asymptotic polarization is proportional to the dimension of the Hilbert space of the scratch qudit (or $n'$ qubits), i.e.
$\epsilon_{\one}^{\infty}\approx md\epsilon_b(=m2^{n'}\epsilon_b)$. As the value of $\epsilon_b$ increases beyond $1/md$, we observe a transition for the asymptotic polarization. This is shown in Fig.\ref{fig:polmax_d}, as a function of the bath polarization for different number of qubits, using eq. (\ref{eq:maxpol}). We can observe the transition noted by \cite{moussa:2005} and \cite{schulman2005physical} at $\epsilon_b \sim 2^{-n}$, for $m=1$, agreeing with simulations. 

In order to see how $\epsilon_{\one}^{\infty}$ approaches 1, we use $\Delta_{max}=1-\epsilon_{\one}^{\infty}$, and eq (\ref{eq:maxpol}). Then,
\begin{equation}
\label{eq:crecerapido}
\Delta_{max}=\frac{2}{e^{{md}\ln\left(\frac{1+\epsilon_b}{1-\epsilon_b}\right)}+1}=\frac{2}{e^{{m2^{n'}}\ln\left(\frac{1+\epsilon_b}{1-\epsilon_b}\right)}+1}.
\end{equation}   
This expression shows that the asymptotic polarization goes to 1 doubly exponentially in the number of qubits $n'$ (or exponential as a function of the size of the Hilbert space $d$). In Fig. \ref{fig:polmax_d}, we show $\epsilon_{\one}^{\infty}$ as a function of $\epsilon_b$ for different values of $d$, with $m=1$.

The asymptotic polarization 
$\epsilon_{\one}^{\infty}$ was obtained assu-ming the system qubits started in the completely mixed state. The same asymptotic polarization would be obtained if we start with a different initial state that nevertheless obeys eq.(\ref{eq:inequal}). Numerical simulation indicates that this could also happens with some initial states not obeying eq.(\ref{eq:inequal}). But we can also find explicit examples of initial states that lead to asymptotic polarizations that are higher than eq.(\ref{eq:maxpol}). As any state can be efficiently maximally randomized, it is always possible to reach the polarization given eq.(\ref{eq:maxpol}) and maybe do better if the initial state is different.

\subsection{Schulman's Physical-Limit Theorem}

The steady state, eq. (\ref{eq:steadystate}), is consistent with the limits of HBAC given by the theorem of Schulman et al. \cite{schulman2007physical}. Their theorem provides an upper bound of the probability of having any basis state, concluding that no heat-bath method can increase that probability from its initial value, $2^{-n}$, to more than $min\{2^{-n}e^{\epsilon 2^{n-1}},1\}$. Where $\epsilon$ is related to the polarization of the heat-bath as $\epsilon_b={\rm tanh}\epsilon$, and $n$ is the total number of qubits ($n=n'+2$: $n'+1$ computational qubits and one reset qubit).

We improved that theorem by finding the correspon-ding exact maximum probability, $p_{max}$. $p_{max}$ is given by the probability of having the basis state $|00...0\rangle$ at the cooling limit: $p_{max}=A_1\left(1+\epsilon_b\right)/2$ (from eq. (\ref{eq:steadystate}) and $\rho=\widetilde\rho_{com}\otimes\rho_{\epsilon_b}$). That expression can by written as a function of $n$ and $\epsilon_b$ as follows $p_{max}=\frac{\epsilon_b}{1-\left(\frac{1-\epsilon_b}{1+\epsilon_b}\right)^{2^{n-1}}}$. 

\begin{figure}[h] 
\center{\includegraphics[width=0.95\linewidth]{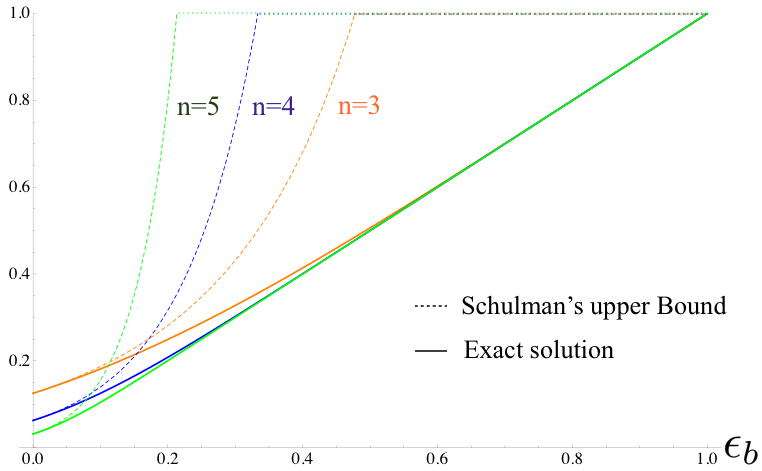}}
\caption{Upper limit of the probability of any basis state for the total $n$ qubit system ($n=n'+2$: $n'+1$ computational qubits and one reset qubit). The dashed line corresponds to the Schulman's upper bound and the thick line to the exact asymptotic probability. Orange for $n=3$, blue for $n=4$, and Green for $n=5$}
\label{fig:coo}
\end{figure}

Fig. \ref{fig:coo} shows both the upper bound proposed by Schulman (dashed lines) and the asymptotic value obtained here (thick lines), for different values of $n$. We can see that the bound is very close to the exact solution for small values of $\epsilon_b$, but differ for large values of $\epsilon_b$. 

\subsection{Number of steps needed to get $\epsilon=\epsilon_{\one}^{\infty}-\delta$} 
We calculated the number of steps required to get a certain polarization for the three qubit case (m=1, d=2). For this, we studied the polarization evolution after each step of the PPA method on the system, starting from the total mixed state. The required quantum circuit to perform the PPA method is shown in Fig. \ref{fig:circuit}.

\begin{figure}[h]
\center{\includegraphics[width=1\linewidth]{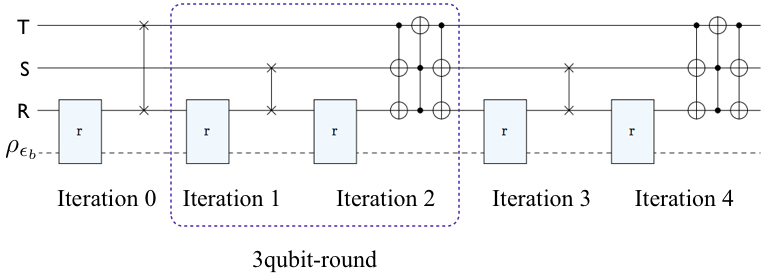}}
\caption{Quantum circuit for the PPA method on a system of three qubits starting in total mixed state. In the circuit diagram, the target, the scratch and the reset qubits are denoted T, S, and R, respectively; the dashed line corresponds to the heat-bath and $r$ stands for the refresh operation. The figure shows only the first five iterations of the circuit (an iteration consists of one refresh step plus one compression step), subsequent iterations are just the repetition of the iterations 1 and 2 (a 3qubit-round).}
\label{fig:circuit}
\end{figure}

Consider that the polarization of the first qubit is $\epsilon^t$ after the $t^{th}$ iteration. Applying two more iterations, which corresponds to the 3qubit-round in Fig. \ref{fig:circuit}, the polarization of the target qubit increases from $\epsilon^t$ to $\epsilon^{t+2}$ as follows: 
\begin{equation}
\label{eq:pol_after}
\epsilon^{t+2}=2ab\epsilon^t+\epsilon_b,
\end{equation} where $a=\frac{1+\epsilon_b}{2}$ and $b=\frac{1-\epsilon_b}{2}$.

Let t starts from 0, then $\epsilon^0=\epsilon_b$ after the first iteration. From eq.(\ref{eq:pol_after}), the polarization after applying $j$ 3qubit-rounds can be written as
\begin{equation}
\label{eq:Steps2}
\epsilon^{t=2j}=\epsilon_{\one}^{\infty}-q^j\left(\epsilon_{\one}^{\infty}-\epsilon_b\right),
\end{equation}
where $q=\frac{1-\epsilon_b^2}{2}$. Using (\ref{eq:maxpol}) with $d=2$, we have that the corresponding asymptotic polarization $\epsilon_{\one}^{\infty}=\frac{2\epsilon_b}{1+\epsilon_b^2}$. 
From this equation we can find the number of steps needed to get to $\epsilon=\epsilon_{\one}^{\infty}-\delta$,
\begin{equation}
\label{eq:Stepstot}
N(\delta,\epsilon_b)=2j= 2\frac{\log\left(\frac{\delta}{\epsilon_{\one}^{\infty}-\epsilon_b}\right)}{\log q}.
\end{equation}

The upper bound on the number of steps required to get polarization $\epsilon_{h,\delta}<\epsilon_{max}$ for the cases of a string of $n$ qubits ($n'=n-2$, $m=1$) is
\begin{equation}
\label{eq:upperbound}
N_{upper-bound}=\prod_{k=1}^{k=[n'/2]}{N(\delta_k,\epsilon_{k}}),
\end{equation} 
where $\epsilon_{max}=\frac{\left(1+\epsilon_b\right)^{d/2}-\left(1-\epsilon_b\right)^{d/2}}{\left(1+\epsilon_b\right)^{d/2}+\left(1-\epsilon_b\right)^{d/2}
}$;
$\epsilon_{k}:=f(\epsilon_{k-1})-\delta_k$;
$\epsilon_{h,\delta}=\epsilon_h$, with $h=[n'/2]$ (the integer part of $n'/2$);
 $f(\epsilon)=\frac{2\epsilon}{1+\epsilon^2}$;
$N(\delta,\epsilon)=2\frac{{\rm log} \left(\frac{\delta}{f(\epsilon_b)-\epsilon_b}\right)}{\rm{log} q}$;
and $\epsilon_0=\epsilon_b$. (More details in Supplemental Material.)

Fig. \ref{fig:Ite_fun_delta_} shows numerical simulations of the number of steps 
as a function of $\delta_{rel}=\frac{\epsilon_{\one}^{\infty}-\epsilon}{\epsilon_{\one}^{\infty}}=\delta/\epsilon_{\one}^{\infty}$.
\begin{figure}[h] 
\center{\includegraphics[width=0.9\linewidth]{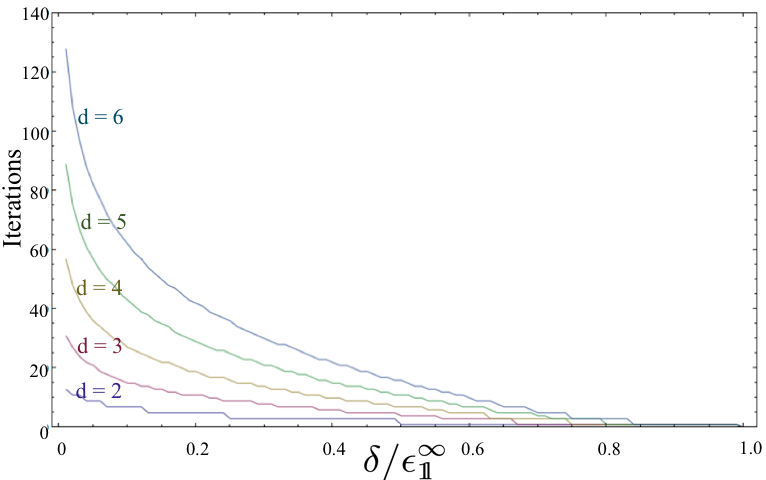}}
\caption{The PPA-steps required to have polarization $\epsilon=\epsilon_{\one}^{\infty}-\delta$ as a function of $\delta/\epsilon_{\one}^{\infty}$, for d=2, 3, 4, 5, and 6.}
\label{fig:Ite_fun_delta_} 
\end{figure}
The simulations are consistent with the upper bound of the number of steps and with the exact solution for the case of three qubits.

\section{Conclusion}

HBAC is a process to purify a number of qubits by removing entropy using extra qubits and contact with a bath. 
We presented an analytical solution for the steady state which corresponds to the cooling limit of a string of qubits starting with the totally mixed state which consists of one qubit with a number of ancilla qubits (or a spin-$l$) and another set of $m$ qubits that can be put into contact with a bath with polarization $\epsilon_b$. From this formula we can understand the transition of behavior of the asymptotic polarization at $1/md$. Below this value, $\epsilon_{\one}^{\infty}\sim md \epsilon_b$ and above it will reach order unity double exponentially with the number of scratch qubits. This behavior will remain true for other initial states as long as they obey conditions (\ref{eq:inequal}). We can think of our derived asymptotic polarization as the minimum polarization limit as it is always possible to efficiently randomized a state so that value can always be asymptotically reached. If conditions (\ref{eq:inequal}) are not obeyed, it may be possible to reach higher polarization. Finally, we obtained the number of steps required to reach a given polarization for a specific number of qubits \footnote{Previously, we reported the formula in eq.(\ref{eq:maxpol}) in a poster and numerics which convincingly demonstrate the obtainable efficiency of HBAC, see arXiv: 1412.6637 (see poster attached).}

\section{Acknowledgements}
The authors would like to thank Xian Ma, Osama Moussa, Om Patange, Tal Mor, Yossi Weinstein, Aharon Brodutch, and Daniel Park for insightful discussions. This work is supported by Industry Canada, the government of Ontario, CIFAR and NSERC. N.A~R.B~is supported by CONACYT-COZCYT and SEP. 

\bibliographystyle{apsrev4-1}
\bibliography{maxhabc}

\newpage

\begin{center}
\large{\bf SUPPLEMENTAL MATERIAL}
\end{center}

Here we explain how to obtain the achievable polarization for heat bath algorithmic cooling (HBAC) when the initial state is totally mixed. First, we give the conditions to have a steady state.
Then, we show that these conditions can be reached asymptotically when we start from the maximally mixed state. From the steady state, we derive the maximum polarization achievable. Furthermore, we explain how
to get the number of steps needed to have a certain polarization
$\epsilon_{\one}^{\infty}-\delta$ (we give the exact solution for $n=3$, and upper bound for $n>3$). (See also [21].) 

\section*{COOLING LIMIT}

In the cooling limit it is not possible to continue extracting entropy from the computational qubits. Thus, the corresponding state, $\rho_{com}$, will not change by applying the compression and refresh steps of HBAC.

The method to find this steady state is to consider the general form of $\rho_{com}$, and apply the two steps of the HBAC method to get $\rho''_{com}$. The conditions for the steady state are given by the equality of these states.

Assume that we start with a system in the totally mixed state. By applying compression and refresh operations, the state remains diagonal. Thus, the state of the whole qubit system, $\rho$, can be completely described by its diagonal elements,
\begin{equation}
\label{eq:dengeneral} \tag{S1}
diag(\rho)= \begin{bmatrix}
 p_1\\
 p_2\\
 . \\
 .\\
 .\\
 p_{D}\\
 \end{bmatrix},
\end{equation}
where $diag(\rho)$ is the vector of the diagonal elements of $\rho$, and D is the dimension of the Hilbert space of the whole string of qubits ($D=2d2^m$).

Applying HBAC, the state evolves through the following two steps:


{\bfseries Entropy Compression Step: $\rho  \xrightarrow{Compress} \rho'=U\rho U^\dagger$.} In the PPA, $U$ sorts the diagonal elements of $\rho$ in decreasing order, giving a $\rho'$ with diagonal elements
\begin{equation}
\label{eq:ordenpi} \tag{S2}
p'_1 \geq  p'_2 \geq ... \geq p'_{D-1} \geq p'_{D}.
\end{equation}

The state of the computational qubits, $\rho'_{com}$, is given by
\begin{equation}
diag(\rho'_{com})=diag({\rm Tr}_m(\rho')):=\begin{bmatrix}
 A_1\\
 A_2\\
 . \\
 .\\
 .\\
 A_{2d}\\
 \end{bmatrix},\tag{S3}
\end{equation}
where ${\rm Tr}_m()$ is the partial trace operation over the $m$ reset qubits, and $A_k= \displaystyle  \sum_{j=j_{k_0}}^{j_{k}} p'_j$, with $j_{k_0}=(k-1)2^m+1$ and $j_{k}=k2^m$. This, with eq. (\ref{eq:ordenpi}), implies that 
\begin{equation}
\label{eq:ordenAi} \tag{S4}
A_1 \geq  A_2 \geq ... \geq A_{2d-1} \geq A_{2d}.
\end{equation} 


{\bfseries Refresh Step: 
$\rho'  \xrightarrow{Refresh}  \rho''=\mathrm{{\rm Tr}}_m\left(\rho'\right)\otimes\rho_{\epsilon_b}^{\otimes m} $,} 
where $\rho_{\epsilon_b}= \frac{1}{2} \begin{pmatrix}
 1+\epsilon_b& 0\\
 0 & 1-\epsilon_b\\
 \end{pmatrix}$ is the state of a qubit with heat-bath polarization $\epsilon_b$. 
 
After these compression and refresh steps, the state of the total qubit system, $\rho''$, will be described by
\begin{equation}
\label{eq:ref3} \tag{S5}
diag\left(\rho''\right)= 
\begin{bmatrix}
 A_1\\
 A_2\\
 . \\
 .\\
 .\\
 A_{2d-1}\\
 A_{2d}\\
 \end{bmatrix} \otimes \frac{1}{2^m}
\begin{pmatrix}
  1+\epsilon_b \\
  1-\epsilon_b\\
 \end{pmatrix}^{\otimes m}.
\end{equation}


In the cooling limit there is no operation that can compress any further the entropy of the computational qubits, or equivalently, the diagonal elements of $\rho''$ are already sorted in decreasing order.


Starting with the simplest case, m=1 (using only one reset qubit), the $diag(\rho'')$ is as follows (from eq.(\ref{eq:ref3})):
\begin{equation}
\label{eq:ref4} \tag{S6}
diag\left(\rho''\right)=
\frac{1}{2} \begin{bmatrix}
 A_1 \left(1+\epsilon_b\right)\\
 A_1 \left(1-\epsilon_b\right)\\
 A_2 \left(1+\epsilon_b\right)\\
 A_2 \left(1-\epsilon_b\right)\\

 . \\
 .\\
 .\\
A_{2d} \left(1+\epsilon_b\right)\\
A_{2d} \left(1-\epsilon_b\right)\\
 \end{bmatrix}.
\end{equation}

If the elements of $\rho''$ are already sorted, it implies that 
\begin{equation}
\label{eq:hold1} \tag{S7}
A_i(1-\epsilon_b)\geq A_{i+1}(1+\epsilon_b), 
\end{equation}
for all $i= 1, 2,..., 2d-1$, which is a condition required for a steady state under the PPA-HBAC.
Note that there are many solutions to this set of equations,
and, not surprisingly, many steady states of HBAC.



Now, we will show that we can reach a steady state if we start from the totally mixed state.

Let $A^t_i$ be the evolution of $A_i$ after $t$ iterations of the PPA-HBAC, with $A^0_i=\frac{1}{2d}$ when the initial state is totally mixed. Interestingly, we have 
\begin{equation}
A^0_i(1-\epsilon_b)\leq A^0_{i+1}(1+\epsilon_b), 
\label{eq:hold1b} \tag{S8}
\end{equation} 
for all $i=1, 2,..., 2d-1$. Note that it is a less than equal sign in distinction from (\ref{eq:hold1}).
We will show that if (\ref{eq:hold1b}) is true at $t=0$, it will be
true for all future steps $t$. Moreover, we will also show that if
(\ref{eq:hold1b}) is obeyed, the rounds of HBAC keep cooling the computational qubits. Thus, the state of the system reaches asymptotically the condition of (\ref{eq:hold1}) with the equality.

We will prove that if we have 
$\frac{A^t_i}{A^t_{i+1}}\leq\frac{1+\epsilon_b}{1-\epsilon_b}$ for all $i=1, 2,..., 2d-1$
at a given moment $t$, then after an iteration of HBAC we will have $\frac{A^{t+1}_i}{A^{t+1}_{i+1}}\leq\frac{1+\epsilon_b}{1-\epsilon_b}$.

Let $\rho^t_{com}$ be the state of the computational qubits after $t$ iterations. Then, the density matrix of the total qubit system state will be given by $\rho^t=\rho^t_{com}\otimes\rho_{\epsilon_b}$, just after a refresh step. Thus, the total state is as follows:

\begin{equation}
\label{colores} \tag{S9}
diag(\rho^{t})=
\begin{bmatrix}
 p^t_1\\
 p^t_2\\
 p^t_3\\
 p^t_4\\
 p^t_5\\
 p^t_6\\
 . \\
 .\\
 .\\
p^t_{2(2d)-1}\\
p^t_{2(2d)}\\
 \end{bmatrix}
 =
\frac{1}{2} \begin{bmatrix}
 {\color{blue} A^t_1 \left(1+\epsilon_b\right)}\\
  {\color{red}  A^t_1 \left(1-\epsilon_b\right)}\\
 {\color{blue} A^t_2 \left(1+\epsilon_b\right)}\\
  {\color{red} A^t_2 \left(1-\epsilon_b\right)}\\ 
   {\color{blue} A^t_3 \left(1+\epsilon_b\right)}\\
  {\color{red} A^t_3 \left(1-\epsilon_b\right)}\\ 
 . \\
 .\\
 .\\
  {\color{blue} A^t_{2d} \left(1+\epsilon_b\right)}\\
  {\color{red} A^t_{2d} \left(1-\epsilon_b\right)}\\ 
 \end{bmatrix}.
\end{equation}

The elements of $\rho^t$ can be written as
\begin{align}
\label{eq:los_rho1} \tag{S10}
&p^t_{2i-1}=A^t_i (1+\epsilon_b)/2, \rm{and}  \\
& p^t_{2i}=A^t_i (1-\epsilon_b)/2,
\label{eq:los_rho}  \tag{S11}
\end{align}
for $i=1,2,...,2d$.

For the next step, we have to compress $\rho^t$ to get $\rho^{t+1}$, i.e. we have to sort the diagonal elements of $\rho^t$ in decreasing order.

Observe that the elements with factor $(1+\epsilon_b)$ (the blue elements
in (\ref{colores})) are already in descending order, since $A^t_1\geq
A^t_2\geq ...\geq A^t_{2d}$. Therefore, during the compression step, these
elements can be moved to different entries of the diagonal matrix
from the initial ones, but they
will have the same order among them (because they are already sorted).
It is similar for the elements with factor $(1-\epsilon_b)$ (the red elements).

Assuming
$\frac{A^t_i}{A^t_{i+1}}\leq\frac{1+\epsilon_b}{1-\epsilon_b}$, as we have in the initial state, implies that the blue elements are going to go up at least one row, except for $A^t_1(1+\epsilon_b)$ which stays in the same position. Similarly, the red elements are going to go down at least one row, except for $A^t_{2d}(1-\epsilon_b)$ which stays in the same position.

Considering this element movement, we can conclude that the elements of $\rho^{t+1}$ will satisfy the following inequalities:
\begin{align}
\label{eq:rho_2i-1} \tag{S12}
&A^t_{i-1}(1-\epsilon_b)/2 \leq p^{(t+1)}_{2i-1}\leq A^t_{i}(1+\epsilon_b)/2, \rm{and} \\
& A^t_{i}(1-\epsilon_b)/2\leq p^{(t+1)}_{2i}\leq A^t_{i+1}(1+\epsilon_b)/2,
\label{eq:rho_2i} \tag{S13}
\end{align}
for $i=2,3,...,2d-1$.
  
 The new computational state, $\rho^{t+1}_{com}={\rm Tr}_m(\rho^{t+1})$, will
 have diagonal elements $A^{t+1}_{i}=p^{(t+1)}_{2i-1}+p^{(t+1)}_{2i}$. From this and (\ref{eq:rho_2i-1})-(\ref{eq:rho_2i}), we have
\begin{align}
\label{eq:At+1} \tag{S14}
&(A^t_{i-1}+A^t_{i})(1-\epsilon_b)/2 \leq A^{(t+1)}_{i}\leq (A^t_{i}+A^t_{i+1})(1+\epsilon_b)/2,
\end{align}
for $i=2,3,...,2d-1$. For the first and last diagonal elements of $\rho_{com}$ ($i=1$ and $i=2d$), we know exactly their corresponding values,
\begin{align}
\label{eq:At+1_1} \tag{S15}
&A^{t+1}_1=(A^t_1+A^t_2)(1+\epsilon_b)/2, \rm{and}\\
&A^{t+1}_{2d}=(A^t_{2d-1}+A^t_{2d})(1-\epsilon_b)/2.
\label{eq:At+1_2} \tag{S16}
\end{align}

These last three equations imply that
$\frac{A^{t+1}_{i}}{A^{t+1}_{i+1}}$ satisfy the following inequality:
\begin{equation}
\label{eq:proof} \tag{S17}
\frac{A^{t+1}_{i}}{A^{t+1}_{i+1}} \leq \frac{A^{t}_{i}(1+\epsilon_b)+A^{t}_{i+1}(1+\epsilon_b)}{A^{t}_{i}(1-\epsilon_b)+A^{t}_{i+1}(1-\epsilon_b)}=\frac{1+\epsilon_b}{1-\epsilon_b},
\end{equation}
for all $i= 1, 2,..., 2d-1$, as we claimed. 

\subsection{Increasing purity}

 We now show that starting in the totally mixed state and applying steps of HBAC, the system will asymptotically go to a state that satisfies the equality in (\ref{eq:hold1}) ([21] gives an alternative argument). To show this, we will prove that the target qubit (the spin$-1/2$) is cooled after each iteration of HBAC, and the reset qubit keeps extracting entropy from the system (cooling the system) after each iteration. All this drives asymptotically the initial state to the steady state.

Consider the state of the system after t iterations, (state of the eq. (\ref{colores})). Then, the reduced density matrix for the target qubit is
\begin{equation}
\label{eq:target} \tag{S18}
diag(\rho^t_{target})=
\begin{bmatrix}
\rho^t_{00_{target}}\\
\rho^t_{11_{target}}\\
\end{bmatrix},
\end{equation}
where $\rho^t_{00_{target}}=\displaystyle \sum_{i=1}^{2d} p^t_i=\sum_{i=1}^{d} A^t_i$, and $ \rho^t_{11_{target}}=1- \rho^t_{00_{target}}$.

Since the compression step reorders the diagonal elements of $\rho^t$ in decreasing order, it is clear that the first $2d$ elements
of the new state, $\rho^{t+1}$, will satisfy 
 $\displaystyle \sum_{i=1}^{2d} p^{t+1}_i \geq \sum_{i=1}^{2d} p^t_i$,
\begin{equation}
\label{eq:purificate} \tag{S19}
 \Longrightarrow \rho^{t+1}_{00_{target}}\geq \rho^{t}_{00_{target}}.
\end{equation}
Therefore, the target qubit is always colder (or remains same) after each iteration of HBAC. 

On the other hand, the reset qubit, which has reduced density matrix $\rho^t_{r}$ when the total system has state $\rho^t$, will be
\begin{equation}
\label{eq:resetqubit} \tag{S20}
diag(\rho^{t+1}_{r})=
\begin{bmatrix}
\rho^{t+1}_{00_{r}}\\
\rho^{t+1}_{11_{r}}\\
\end{bmatrix},
\end{equation}
where $\rho^{t+1}_{00_{r}}=\displaystyle \sum_{i=1}^{2d} p^{t+1}_{2i-1}$. This equation, with (\ref{eq:rho_2i-1}) and (\ref{eq:los_rho1}), gives
\begin{equation}
\label{eq:desi_resetqubit} \tag{S21}
\rho^{t+1}_{00_{r}}=\sum_{i=1}^{2d} p^{t+1}_{2i-1} {\leq} \sum_{i=1}^{2d} A^{t}_{i} (1+\epsilon_b)/2= (1+\epsilon_b)/2.
\end{equation}
Therefore, the reset qubit will always be hotter than the
bath after the compression step of HBAC as long as we do not reach the equality. This implies that the reset qubit always extracts entropy from the total system when it is brought
into contact with the heat bath. The system is cooled in every
iteration of the refresh step, with a smaller and smaller amount of entropy extracted, going asymptotically the cooling limit.


The two elements above show that, starting from the totally mixed state, we will converge to the equality 
of (\ref{eq:hold1}). At this limit, the steady state of the
computational qubits should have elements which satisfy
\begin{equation}
\label{eq:condicionfinal} \tag{S22}
\frac{A^\infty_{i+1}}{ A^\infty_i} = 
\frac{1-\epsilon_b}{1+\epsilon_b} \equiv Q.
\end{equation} 
 
Using (\ref{eq:condicionfinal}) and $Tr(\rho_{com})=1$, it is possible to find the exact solution of each $A^\infty_i$:
\begin{equation}
\label{eq:As} \tag{S23}
A^\infty_{i}=\frac{1-Q}{1-Q^{2d}}Q^{i-1},
 \end{equation}
and therefore the analytical solution of the steady state of the computational qubits will be
\begin{equation}
\label{eq:steadystateeee} \tag{S24}
diag\left(\rho^\infty_{com} \right)= 
A^\infty_1\begin{bmatrix}
 1\\
 Q\\
 Q^2\\
 .\\
 .\\
 .\\
 Q^{2d-1}\\
 \end{bmatrix}.
 \end{equation}

\subsection{Asymptotic Polarization of the target qubit for one and multiple reset qubits}

Using eq.(\ref{eq:steadystateeee}), the reduced density matrix of the target qubit in the cooling limit is given by 
\begin{equation}
\label{eq:densiii1} \tag{S25}
diag(\rho^\infty_{target})
 =A^\infty_1 \sum_{i=0}^{d-1}Q^i 
\begin{bmatrix}
 1 \\
Q^{d} \\
 \end{bmatrix}
=
\frac{1}{2}
 \begin{bmatrix}
 1+\epsilon^\infty_{\one} \\
1-\epsilon^\infty_{\one}\\
 \end{bmatrix},
\end{equation}
where $\epsilon^\infty_{\one}$ is the asymptotic polarization of the target qubit when we start with the maximally mixed state. 

From this equation we can derive:
\begin{equation}
\label{eq:maxpol_d} \tag{S26}
\epsilon^\infty_{\one}=\frac{\left(1+\epsilon_b\right)^{d}-\left(1-\epsilon_b\right)^{d}}{\left(1+\epsilon_b\right)^{d}+\left(1-\epsilon_b\right)^{d}
},
\end{equation}
where $d$ is the dimension of the Hilbert space of the scratch qudit
($d=2l+1$ if we use a spin$-l$, or $d=2^{n'}$ if we use a string of $n'$ qubits).

Now, if we generalize to the case $m>1$, we have that the state of the $m$ reset qubits is given by

\begin{equation} \tag{S27}
diag(\rho_{\epsilon_b}^{\otimes m})=
\begin{pmatrix}
  1+\epsilon \\
  1-\epsilon\\
 \end{pmatrix}^{\otimes m}=
 \begin{pmatrix}
  (1+\epsilon)^m\\
  . \\
 .\\
 .\\
  (1-\epsilon)^m\\
 \end{pmatrix},
\end{equation}
where $(1+\epsilon_b)^m$ is the biggest element, and   $(1-\epsilon_b)^m$ the smallest one, which correspond to the first entry and the last entry, respectively. Observe that in general the diagonal elements of $\rho_{\epsilon_b}^{\otimes m}$ are not in decreasing order.

From eq. (\ref{eq:ref3}), $\rho''$ is as follows:
\[diag(\rho'')=
\begin{tikzpicture}[baseline=-\the\dimexpr\fontdimen22\textfont2\relax ]
\matrix (m)[matrix of math nodes,left delimiter=(,right delimiter=)]
{
A_1 \left(1+\epsilon_b\right)^m\\
  . \\
 .\\
 .\\
 A_1 \left(1-\epsilon_b\right)^m\\
 A_2 \left(1+\epsilon_b\right)^m\\
 . \\
 .\\
 .\\
 A_2 \left(1-\epsilon_b\right)^m\\
 . \\
 .\\
 .\\
 A_i \left(1+\epsilon_b\right)^m\\
 . \\
 .\\
 .\\
 A_i \left(1-\epsilon_b\right)^m\\
  A_{i+1} \left(1+\epsilon_b\right)^m\\
 . \\
 .\\
 .\\
 A_{i+1} \left(1-\epsilon_b\right)^m\\
 . \\
 .\\
 .\\
A_{2d} \left(1+\epsilon_b\right)^m\\
 . \\
 .\\
 .\\
A_{2d} \left(1-\epsilon_b\right)^m\\
};.

\begin{pgfonlayer}{myback}
\fhighlight[white]{m-1-1}{m-5-1}
\fhighlight[white]{m-6-1}{m-10-1}
\fhighlight[white]{m-14-1}{m-18-1}
\fhighlight[white]{m-19-1}{m-23-1}
\fhighlight[white]{m-27-1}{m-31-1}
\end{pgfonlayer}
\end{tikzpicture}
\]

First, notice that any swap between two elements within the same box
(which has the same factor $A_i$) will not improve the entropy
compression on the computational qubits state. The reason is once the
reset qubits are traced out, the permutation inside the same box contributes to the sum of the
probabilities corresponding to same basis state of the computational
qubits that they contributed before the compression.

Then, we are just interested in permuting elements to a different box from where they were previously, in particular the biggest element or smallest element of each box (to have the maximum entropy compression). At the cooling limit, there is no operation that can improve the compression, or equivalently, the elements (just taking the largest and smallest of each box) are already sorted.

Following the same reasoning to the case when $m=1$, the steady state should have elements which hold:
\begin{equation}
\label{eq:hold2} \tag{S28}
A ^{\infty}_i(1-\epsilon_b)^m\geq A ^{\infty}_{i+1}(1+\epsilon_b)^m.
\end{equation}

Moreover, similarly to the case of $m=1$, the inequality $\frac{A_i}{A_{i+1}}\leq \frac{(1+\epsilon_b)^m}{(1-\epsilon_b)^m}$ cannot be inverted by applying the steps of HBAC. Therefore, if we start with a totally mixed state (which holds the last inequality mentioned), the steady state should have elements which hold
\begin{equation}
\label{eq:hold3} \tag{S29}
A ^{\infty}_i(1-\epsilon_b)^m=A ^{\infty}_{i+1}(1+\epsilon_b)^m.
\end{equation}

Then, the analytical solution of the steady state of the computational
qubits will be
\begin{equation}
\label{eq:steadystateeee2} \tag{S30}
diag\left(\rho^\infty_{com} \right)= 
A^\infty_1\begin{bmatrix}
 1\\
 Q^m\\
 Q^{2m}\\
 .\\
 .\\
 .\\
 Q^{(2d-1)m}\\
 \end{bmatrix}.
 \end{equation}
Similarly, the maximum achievable polarization using $m$ reset qubits will be
\begin{equation}
\label{eq:maxpolS} \tag{S31}
\epsilon_{\one}^{\infty}=
\frac{\left(1+\epsilon_b\right)^{md}-\left(1-\epsilon_b\right)^{md}}
         {\left(1+\epsilon_b\right)^{md}+\left(1-\epsilon_b\right)^{md}}.
\end{equation}

Note that a similar polarization would be obtained if we start with a different initial state but which obeys eq.(\ref{eq:hold1b}). Numerical simulation indicate that this could also happens with some
initial states not obeying eq.(\ref{eq:hold1b}). Finally, we can give explicit examples of initial states 
that lead to an asymptotic polarization higher than eq.(\ref{eq:maxpol}).

\subsection{Temperature in the cooling limit}

The state of the heat-bath in thermal equilibrium, temperature $T_b$, is given by 

$\rho_{b}= \frac{1}{e^{\Delta E_b/2kT_b}+e^{-\Delta E_b/2kT_b}} \begin{pmatrix}
 e^{\Delta E_b/2kT_b} & 0\\
 0 & e^{-\Delta E_b/2kT_b}\\
 \end{pmatrix}$,
where $\Delta E_b$ is the energy gap between the two energy levels of a qubit from the bath.

Then, the heat-bath polarization corresponds to $\epsilon_b=\rm tanh\left(\frac{\Delta E_b}{2kT_b}\right)$, or equivalently, 
\begin{equation} \tag{S32}
\displaystyle \frac{\Delta E_b}{2kT_b}=\frac{1}{2} {\rm log} \left[\frac{1+\epsilon_b}{1-\epsilon_b}\right].
\end{equation}

Similarly for the target qubit in the steady state at temperature $T_{steady}$, we will have $\displaystyle \frac{\Delta E_t}{2kT_{steady}}=\frac{1}{2} {\rm log} \left[\frac{1+\epsilon^\infty_{\one}}{1-\epsilon^\infty_{\one}}\right]$, where $\Delta E_t$ is the energy gap of the two energy levels of the target qubit. From this and eq.(\ref{eq:maxpol}), we can obtain the temperature in the cooling limit,

\begin{equation} \tag{S33}
\displaystyle T_{steady}=\left(\frac{1}{md}\right) T_{b}\left(\frac{\Delta E_t}{ \Delta E_b}\right),
\end{equation}
$d=2^{n'}$ when the scratch qudit is a string of $n'$ qubits ($n'+1$ computational qubits).

The PPA-HBAC method is in line with the third law of thermodynamics, which says that ``it is impossible by any procedure, no matter how idealized, to reduce any assembly to absolute zero temperature in a finite number of operations" [PRE 85, 061126 (2012)], (see also arXiv: 1412.3828). Indeed, the evolution of the state of the system goes asymptotically to a steady state, which has non zero temperature for a finite number of qubits. The limit when the temperature is exactly zero corresponds to the case of having an infinite number of qubits. Since the number of gates needed grows with the number of qubits, the operations required to achieve temperature zero will be infinite.

Although the algorithm keeps cooling the target qubit at each time, it does so with a smaller and smaller amount of entropy extracted, asymptotically reaching the steady state of non-zero temperature. This is in agreement with the third law of thermodynamics.

\subsection{Polarization of different computational qubits}

Consider the case of having a string of $n'$ qubits as scratch
qubit. Let's label the qubits from right to left, as it is shown in Fig. 1 in the paper.

We can obtain the polarization of each qubit
from the steady state (\ref{eq:steadystateeee}). We already showed how to get the
polarization of the target qubit. If we trace out the target qubit
from the computational qubits, we
can repeat the same calculations to get the polarization of the
neighbor qubit in the string (which is labeled as qubit $n'$) since
this qubit will be now the first from the left.

The state of the computational qubits without the target qubit is
\begin{equation} \tag{S34}
diag(\rho ^{\infty}_{\bar target}) ={\rm Tr}_{target} (\rho ^{\infty}_{com} )=\begin{bmatrix} 
 A ^{\infty}_1+ A^{\infty}_{d+1} \\
 A ^{\infty}_2+ A^{\infty}_{d+2} \\
 . \\
 .\\
 .\\
 A^{\infty}_d+ A^{\infty}_{2d} \\
 \end{bmatrix} .
\end{equation}

Let $B_i$ be the $ith$ element of the $diag(\rho^{\infty}_{\bar target})$, i.e. $B_i=A^{\infty}_i+A^{\infty}_{d+i}$. From eq. (\ref{eq:As}),
$B_i=A^{\infty}_1Q^{i-1}+A^{\infty}_1Q^{d+i-1}=A^{\infty}_1(1+Q^d)Q^{i-1}$. Thus,
$B_i=kQ^{i-1}$, where $k=A^{\infty}_1(1+Q^d)$. Comparing $B_i$ with
eq(\ref{eq:As}), we see that this state has the same form of the
state eq. (\ref{eq:steadystateeee}), but with Hilbert space dimension
$d/2$. Thus, the asymptotic polarization of the $n' th$ qubit is
\begin{equation}
\label{eq:maxpol_qubitn} \tag{S35}
\epsilon_{max}^{(n^{'})}=\frac{\left(1+\epsilon_b\right)^{m
    d/2}-\left(1-\epsilon_b\right)^{m
    d/2}}{\left(1+\epsilon_b\right)^{m
    d/2}+\left(1-\epsilon_b\right)^{m d/2}}
\end{equation}
where $d=2^{n'}$.

Similarly, we can get the polarization of the $(n'-1)^{th}$ qubit,
and so on. Then, the polarization of the $j^{th}$ qubit will be
\begin{equation}
\label{eq:maxpol_qubitn} \tag{S36}
\epsilon^{(j)}_{max}=\frac{\left(1+\epsilon_b\right)^{m 2^{j-1}}-\left(1-\epsilon_b\right) ^{m 2^{j-1}}}{\left(1+\epsilon_b\right) ^{m 2^{j-1}}+\left(1-\epsilon_b\right) ^{m 2^{j-1}}
}.
\end{equation}

\subsection{Number of steps needed to get $\epsilon=\epsilon_{\one}^{\infty}-\delta$ } 

\subsubsection{Analytical result for a string of three qubits (m=1, d=2).}

The quantum circuit required to perform the PPA-HBAC on three qubits
initially in the total mixed state is showed in Fig.\ref{fig:circuit}. This circuit shows the operations required for the
first five iterations (each iteration consists of a refresh step and
an entropy compression step). Subsequent iterations gates are the
alternate repetition of the second and third iterations gates in 
Fig.\ref{fig:circuit}. The application of those two iterations will be referred as a 3qubit-round.

\begin{figure}[h] 
\center{\includegraphics[width=1\linewidth]{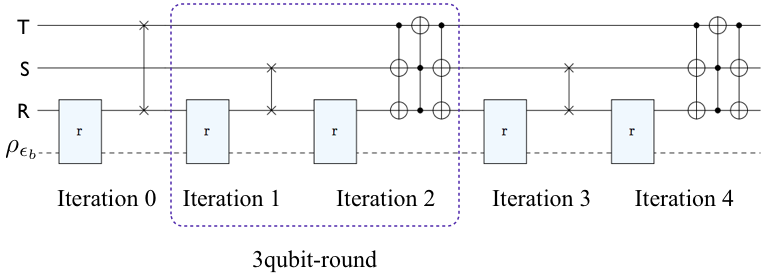}}
\caption{Quantum circuit for the PPA method on a system of three qubits starting in the total mixed state. In the circuit diagram, the target, the scratch and the reset qubits are denoted T, S, and R, respectively; the dashed line corresponds to the heat-bath and $r$ stands for the refresh operation. The figure shows only the first five iterations of the circuit (an iteration consists of one refresh step plus one compression step), subsequent iterations are just the repetition of iterations 1 and 2 (a 3qubit-round).}
\label{fig:circuit} 
\end{figure}

In order to know the effect of one 3qubit-round on the system, consider the state of the computational qubits at a given moment,
\begin{equation}
\label{eq:rho_comp_3q} \tag{S37}
diag(\rho^t_{com})=
\begin{bmatrix}
 A^t_1\\
 A^t_2\\
 A^t_3\\
 A^t_4\\
 \end{bmatrix},
\end{equation}
and the total system as $\rho^t= \rho^t_{com}\otimes\rho_{\epsilon_b}$. The polarization of the target qubit, $\epsilon^t$, can be obtained from its reduced density matrix, 
$diag(\rho^t_{target})=
\begin{bmatrix}
 A^t_1+A^t_2\\
 A^t_3+A^t_4\\
 \end{bmatrix}=\frac{1}{2}
 \begin{bmatrix}
 1+\epsilon^t\\
 1-\epsilon^t\\
 \end{bmatrix}$ 
\begin{equation}
\label{eq:pol_t} \tag{S38}
 \Longrightarrow 
\epsilon^t=2(A^t_1+A^t_2)-1.
\end{equation}

\begin{figure}[h]
\center{\includegraphics[width=1\linewidth]{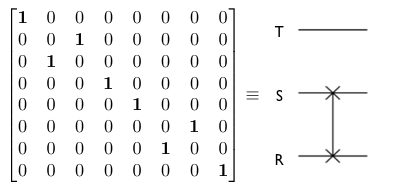}}
\caption{\small{Matrix and circuit symbol corresponding to the entropy
    compression step of iteration 1. This gate swaps the scratch
    qubit and the reset qubit.}}
\label{fig:u2}
\end{figure}

In the first iteration of the 3qubit-round, the compression gate swaps the
scratch qubit and the reset qubit. This swap can be performed by
applying the unitary matrix shown in Fig.\ref{fig:u2}, thus

$diag(\rho^t)= \frac{1}{2}
\begin{bmatrix}
 A^t_1\left(1+\epsilon_b\right)\\
 A^t_1\left(1-\epsilon_b\right)\\ 
 A^t_2\left(1+\epsilon_b\right)\\
  A^t_2\left(1-\epsilon_b\right)\\
   A^t_3\left(1+\epsilon_b\right)\\
 A^t_3\left(1-\epsilon_b\right)\\ 
 A^t_4\left(1+\epsilon_b\right)\\
  A^t_4\left(1-\epsilon_b\right)\\
 \end{bmatrix}
 \Longrightarrow 
  \frac{1}{2}
\begin{bmatrix}
 A^t_1\left(1+\epsilon_b\right)\\
  A^t_2\left(1+\epsilon_b\right)\\
  A^t_1\left(1-\epsilon_b\right)\\ 
  A^t_2\left(1-\epsilon_b\right)\\
   A^t_3\left(1+\epsilon_b\right)\\
 A^t_4\left(1+\epsilon_b\right)\\
  A^t_3\left(1-\epsilon_b\right)\\ 
  A^t_4\left(1-\epsilon_b\right)\\
 \end{bmatrix}.$

Then, the density matrix of the computational qubits after the first iteration of the 3qubit-round is 
\begin{equation}
\label{eq:afterIte2} \tag{S39}
  diag(\rho^{t+1}_{com})
  =\frac{1}{2}
\begin{bmatrix}
 \left( A^t_1+A^t_2 \right) \left(1+\epsilon_b\right)\\
 \left( A^t_1+A^t_2 \right) \left(1-\epsilon_b\right)\\
 \left( A^t_3+A^t_4 \right) \left(1+\epsilon_b\right)\\
 \left( A^t_3+A^t_4 \right) \left(1-\epsilon_b\right)\\
 \end{bmatrix}.
\end{equation} 
 
\begin{figure}[h] 
\center{\includegraphics[width=1\linewidth]{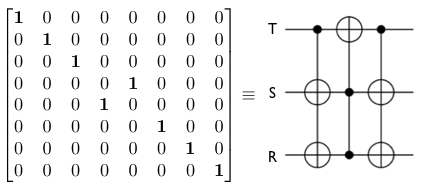}}
\caption{\small{Matrix and circuit symbol corresponding to the entropy
    compression of iteration 2.}}
\label{fig:u3}
\end{figure}
 
In the second iteration of the 3qubit-round, the compression step is
performed by applying the unitary matrix shown in Fig.\ref{fig:u3}. In this step we obtain $\rho^{t+2}$,

$diag(\rho^{t+2})=\frac{1}{4}
\begin{bmatrix}
 \left( A^t_1+A^t_2 \right) \left(1+\epsilon_b\right)^2\\
  \left( A^t_1+A^t_2 \right) \left(1+\epsilon_b\right)\left(1-\epsilon_b\right)\\
 \left( A^t_1+A^t_2 \right) \left(1-\epsilon_b\right)\left(1+\epsilon_b\right)\\
 \left( A^t_3+A^t_4 \right) \left(1+\epsilon_b\right)^2\\
 \left( A^t_1+A^t_2 \right) \left(1-\epsilon_b\right)^2\\
  \left( A^t_3+A^t_4 \right) \left(1+\epsilon_b\right)\left(1-\epsilon_b\right)\\
 \left( A^t_3+A^t_4 \right) \left(1-\epsilon_b\right)\left(1+\epsilon_b\right)\\
  \left( A^t_3+A^t_4 \right) \left(1-\epsilon_b\right)^2\\
 \end{bmatrix}.$

From this state, with the normalization property of the density matrix and (\ref{eq:pol_t}), we can obtain the new polarization of the target qubit,
\begin{equation}
\label{eq:pol_t+2} \tag{S40}
\epsilon^{t+2}=2ab\epsilon^t+\epsilon_b,
\end{equation} 
where $a=\frac{1+\epsilon_b}{2}$ and $b=\frac{1-\epsilon_b}{2}$.

Let $t=0$ (just after the iteration 0 which swaps the target qubit and
the reset qubit, Fig.\ref{fig:circuit}), then the polarization of the target qubit at that moment will be $\epsilon^0=\epsilon_b$. From eq.(\ref{eq:pol_t+2}), we can get the exact polarization after each 3qubit-round, i.e. every two iterations,

\begin{equation}
\label{eq:Steps} \tag{S41}
\epsilon^{t=2j}=\frac{2\epsilon_b}{1+\epsilon_b^2}-q^j\left(\frac{2\epsilon_b}{1+\epsilon_b^2}-\epsilon_0\right),
\end{equation}
where $q=\frac{1-\epsilon_b^2}{2}$. From (\ref{eq:maxpol}), the
asymptotic polarization for this case is
$\epsilon^\infty_{\one}=\frac{2\epsilon_b}{1+\epsilon_b^2}$, thus
eq.(\ref{eq:Steps}) can be written as
\begin{equation}
\label{eq:Steps2} \tag{S42}
\epsilon^{t=2j}=\epsilon^\infty_{\one}-q^j\left(\epsilon^\infty_{\one}-\epsilon_b\right).
\end{equation}
Since $q<1$, $\epsilon^t\to\epsilon^\infty_{\one}$ when we increase $j$. 

We can use (\ref{eq:Steps2}) to know the number of rounds $t$ needed to
achieve polarization
$\epsilon^\infty_{\one}-\delta$. From Eq. (\ref{eq:Steps2}), we have
$\delta=q^j\left(\epsilon^\infty_{\one}-\epsilon_b\right)$, then the number of rounds required will be
\begin{equation}
\label{eq:Stepstot} \tag{S43}
N(\delta,\epsilon_b):=t=2\frac{\rm{log}\left(\frac{\delta}{\epsilon^\infty_{\one}-\epsilon_b}\right)}{\rm{log}q},
\end{equation}
to get polarization
\begin{equation}
\label{eq:pol_delta} \tag{S44}
\epsilon_\delta (\epsilon_b,\delta):=\epsilon^\infty_{\one}-\delta=\frac{2\epsilon_b}{1+\epsilon^2_b}-\delta.
\end{equation}

\subsubsection{Numerical results}

Let
$\delta_{rel}=\frac{\epsilon_{\one}^{\infty}-\epsilon}{\epsilon_{\one}^{\infty}}=\delta/\epsilon_{\one}^{\infty}$. Fig.
S\ref{fig:Ite_fun_delta_}
shows simulations of the number of refresh steps needed to achieve a polarization $\epsilon=\epsilon_{\one}^{\infty}\left(1-\delta_{rel}\right)$ as function of $\delta_{rel}$ for different values of d.
The exact solution of number of steps needed for the
3 qubit case is
consistent with the results from the simulations.

\begin{figure}[h] 
\center{\includegraphics[width=0.9\linewidth]{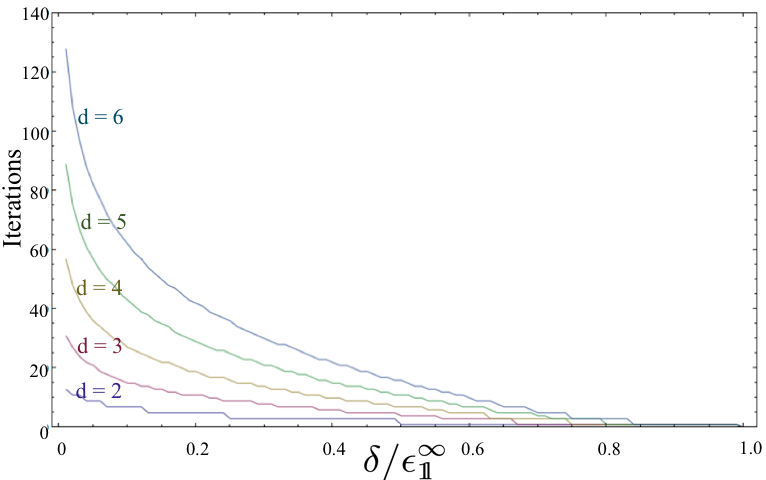}}
\caption{Number of iterations needed to achieve polarization
  $\epsilon=\epsilon_{\one}^{\infty}-\delta$ as a function of $\delta/\epsilon_{\one}^{\infty}$, for d=2, 3, 4, 5, and 6.}
\label{fig:Ite_fun_delta_}
\end{figure}

\subsubsection{Upper bound of the number of steps to get a certain polarization, for n qubits}

Consider a string of $n'+1$ computational qubits, numbered as in Fig. 1 in the paper, and one reset qubit, all starting in totally mixed state.
Applying the compression for three qubits, using the reset qubit and
qubit 1 to cool qubit 2, we can increase the
polarization of qubit 2 to $\epsilon_1=\epsilon_\delta(\epsilon_b,\delta)$ in $N_1=N(\delta,\epsilon_b)$ steps, from (\ref{eq:Stepstot}) and (\ref{eq:pol_delta}). 

After this preparation of qubit 2, we can swap it with qubit 3,
and then prepare again qubit 2. We can apply again the compression
for three qubits, but now using qubits 2 and 3 to cool qubit 4. In
this case, we will need $N_2=N(\delta,\epsilon_1)\cdot N_1$ number of
steps to get polarization
$\epsilon_2=\epsilon_\delta(\epsilon_1,\delta)$ on qubit 4.

We can iterate this idea to use qubit 4 and qubit 5 to cool qubit 6, getting that we need $N_3=N(\delta,\epsilon_2)\cdot N_2$ number of steps to achieve polarization $\epsilon_3=\epsilon_\delta(\epsilon_2,\delta)$, and so on.

Since this is not the optimal compression, this number of iterations
gives an upper bound of the optimal number of steps using the PPA to achieve polarization
$\epsilon<\epsilon_{max}$ on the target qubit, where $\epsilon_{max}=\epsilon^\infty_{\one}=\frac{\left(1+\epsilon_b\right)^{d/2}-\left(1-\epsilon_b\right)^{d/2}}{\left(1+\epsilon_b\right)^{d/2}+\left(1-\epsilon_b\right)^{d/2}
}$, and $\epsilon=\epsilon_\delta(\epsilon_{h-1},\delta)$ with
$\epsilon_0=\epsilon_b$, and $h=[n'/2]$ (the integer part of $n'/2$). The upper bound is
$N_{upper-bound}=\displaystyle \prod_{k=1}^{k=[n'/2]}{N(\delta,\epsilon_{k}})$.

\begin{figure}[h] 
\center{\includegraphics[width=2\linewidth]{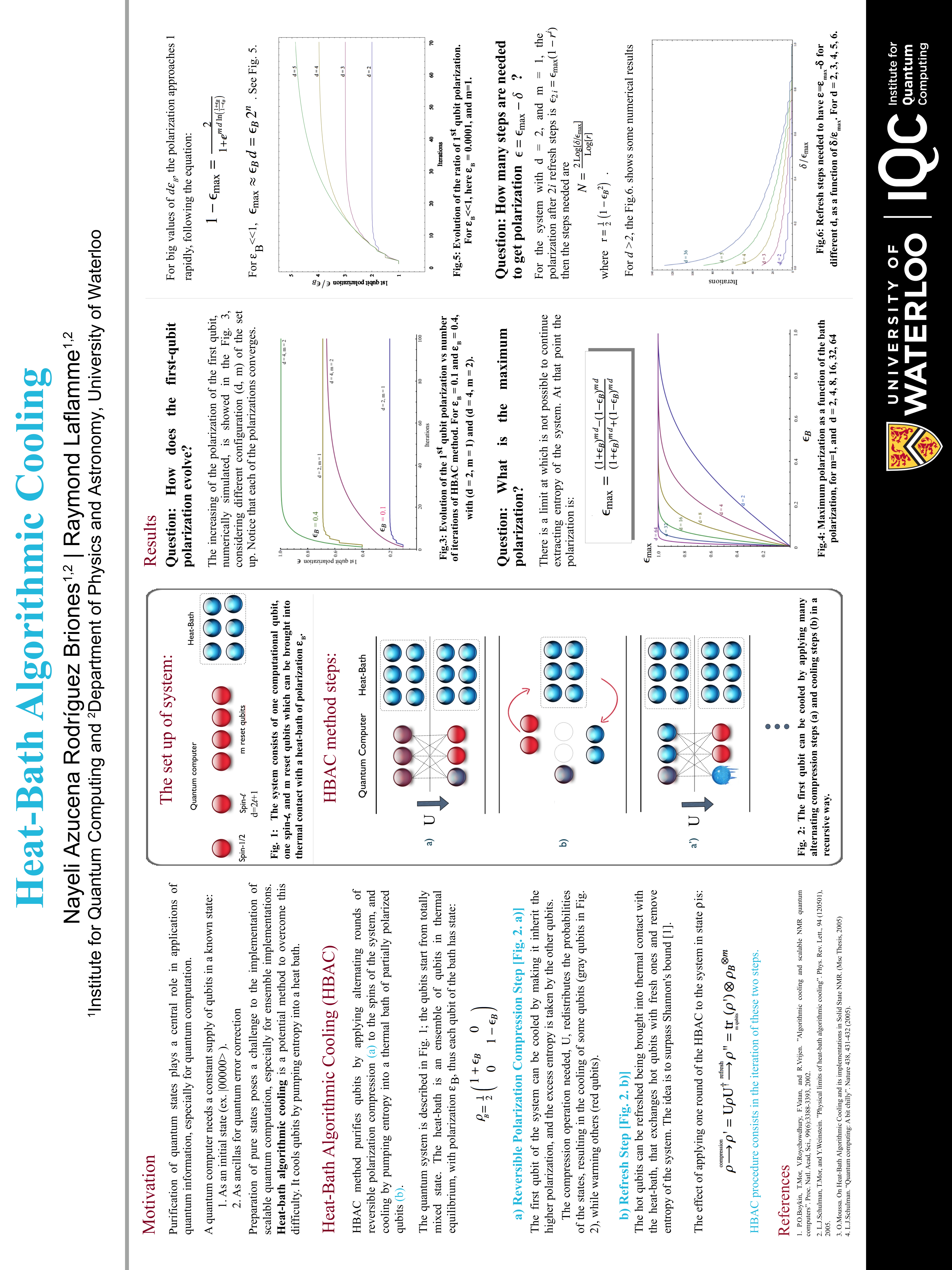}}
\caption{This work was presented on a poster at the Institute for Quantum Computing (March 27, 2014) and at the $14^{th }$ Annual Canadian Summer School on Quantum Information at the Univ. of Guelph (June 16-20, 2014).}
\label{poster}
\end{figure}

\end{document}